\documentclass{aa520}
\usepackage{txfonts}
\newcommand{\be}{\begin{equation}}
\newcommand{\ee}{\end{equation}}
\newcommand{\src}{XTE\,J1946+274}
\newcommand{\srcgro}{GRO\,J1944+26}

\newcommand{\BSAX}{{\em Beppo\/}SAX}
\newcommand{\RXTE}{{\em Rossi }XTE}
\newcommand{\bc}{\begin{center}}
\newcommand{\ec}{\end{center}}

\def\farcs{\hbox{$.\!\!^{\prime\prime}$}}

\input psfig.sty
\begin{document}
\author{F. Verrecchia$^{1,2}$, G.L. Israel$^{3,}$\thanks{Affiliated to
I.C.R.A.}, I. Negueruela$^{4}$, S. Covino$^5$, V.F. Polcaro$^6$, J.S.
Clark$^{7,8}$, I.A.
Steele$^{9}$, R. Gualandi$^{10}$, R. Speziali$^3$ and L.
Stella$^{3,\star}$}

\title{The identification of the optical/IR counterpart of the 15.8-s
transient X--ray pulsar \src}

\institute{
Dipartimento di Fisica, Universit\`a degli Studi ``La Sapienza'',
Piazza
A. Moro 5, I--00185 Roma, Italy
\and
ASI Science Data Center, ESA-Esrin, Via Galileo Galilei, I--00044
Frascati, Rome, Italy \and INAF
-- Osservatorio Astronomico di Roma, Via Frascati 33, I--00040
Monteporzio Catone, Italy \and
Observatoire Astronomique de Strasbourg, rue de l'Universit\'e 11,
F67000 Strasbourg , France \and
INAF -- Osservatorio Astronomico di Brera, Via E. Bianchi 46, I--23807
Merate, Italy \and Istituto
di Astrofisica Spaziale, CNR, Area della Ricerca di Roma Tor Vergata,
Via Fosso del Cavaliere,
I--00133 Roma, Italy \and Astronomy Centre, CPES, University of Sussex,
Brighton BN1 9QH, UK \and
Dept. of Physics \& Astronomy, University College London, London WC1E
6BT, UK \and Astrophysics
Research Institute, Liverpool John Moores University, Liverpool CH41
1LD, UK \and INAF --
Osservatorio Astronomico di Bologna, Via Ranzani, 1, I--40127, Bologna,
Italy }

\date{Received: 15 May 2002 / Accepted: 24 July 2002}
\offprints{verrecchia@mporzio.astro.it}
\authorrunning{Verrecchia et al.}
\titlerunning{Identification of optical/IR counterpart of \src}

\abstract{We report on the discovery of the optical/IR counterpart of
the 15.8\,s transient X--ray
pulsar \src.  We re--analysed archival \BSAX\ observations of \src\,
obtaining a new refined
position (a circle with 22\arcsec\ radius at 90\% confidence level).
Based on this new position we
carried out optical and infra--red (IR) follow--up observations. Within
the new error circle we
found a relatively optical faint ($B$=$18.6$) IR bright ($H$=$12.1$)
early type reddened star ($V$--$R$=$1.6$). The optical spectra show 
strong H$\alpha$ and H$\beta$ emission lines. The
IR photometric observations of the field confirm the presence of an IR
excess for the H$\alpha$--emitting star ($K$=$11.6$, $J$--$H$=$0.6$) which 
is likely surrounded by a circumstellar envelope. Spectroscopic and photometric 
data indicate a B0--1V--IVe spectral--type
star located at a distance of 8--10\,kpc and confirm the
Be--star/X--ray binary nature of \src.
\keywords{stars: individual: -- \src; \srcgro\ -- binaries: general --
stars: pulsars: general --
stars: emission--line, Be -- X--rays: stars } } \maketitle

\section{Introduction}

High Mass X--ray Binary systems (HMXRBs) hosting a neutron star (NS)
are a class of intrinsically
variable and bright X--ray sources whose luminosity is powered by
accretion from a massive
companion, often a B--emission spectral--type star (BeXRBs; White et
al.  1995; Coe 2000).

BeXRBs usually show in the X--ray emission three typical behaviours
(sometimes more than one
within the same source): (i) bright X--ray transients characterized by
giant outburst with
luminosity up to $L_x$\,=\,$10^{38}$\,{\rm erg\,s}$^{-1}$ (Type II;
Stella et al. 1986),
unrelated to the orbital phase and with high spin--up rates, (ii)
transients characterized by
periodic outbursts of relatively high luminosity
($L_x$\,$\simeq$\,$10^{36}$--$10^{37}$\,{\rm
erg\,s}$^{-1}$; Type I) generally occurring at periastron passage of
the NS, and (iii) sources
which do not display outbursts, but are moderately variable (of a
factor up to 10--100) and show a
low luminosity ($L_x$\,$\leq$\,$10^{36}$\,{\rm erg\,s}$^{-1}$) pulsed
persistent emission (Negueruela 1998).

Be stars are early type stars, with mass in the range of
2--20\,$M_{\odot}$ and luminosity class
III to V, which often display Balmer lines in emission in their optical
spectra. These stars are
characterized by high rotational velocity (up to $\sim$70\% of their
break--up velocity), and by
phenomena of equatorial mass ejection at irregular times that give rise
to a rotational--dominated
quasi--Keplerian decretion disc (Hanuschik 1996; Porter 1999; Okazaki
2001; Okazaki \& Negueruela 2001). The widely accepted empirical model 
interprets the X--ray emission from these systems, in
the three mentioned behaviours, as due to sporadic mass accretion of
the NS during the expansion
phase of the Be disc, or to periodic close encounters of the NS in
highly eccentric orbit with the
dense circumstellar envelope around the Be star.

At optical wavelengths these stars are difficult to classify owing to
the presence of emissions
lines produced by the circumstellar envelope and to the high absorption
column that affect systems lying in the Galactic plane.
  So far only about
$\sim$\,30 optical counterparts of BeXRB's have been discovered out of
the $>$\,100 known and
10$^4$--10$^5$ expected (Nelson et al. 1993 \& 1995; Covino et al.
2000; Covino et al. 2001;
Israel et al. 2000a; Israel et al. 2000b; Israel et al. 2001;
Chakrabarty et al. 2002).

The hard X--ray transient source \src\ was discovered with the All Sky
Monitor (ASM) on board the
\RXTE\ satellite during a scan of the Vul--Cyg region on 1998 September
5 (Smith \& Takeshima
1998). \src\ lies inside the error box of the Ariel V transient
3A\,1942+274 discovered in the
1976 (Warwick et al. 1981). Campana et al. (1999) have
estimated a chance probability
of $\sim$\,7\,\% that \src\ and 3A\,1942+274 are not the same source
(assuming $\sim$\,1000 hard
X--ray transients in the Galaxy). The observed flux raised from
$\sim$\,13\,{\rm mCrab}
(2--12\,{\rm keV}) on September 5 to $\sim$\,60\,{\rm mCrab} on
September 15, to $\sim$\,110\,{\rm
mCrab} on September 16 (in 2--60\,{\rm keV}; Smith \& Takeshima, 1998).
Further observations
carried out by BATSE (Wilson et al. 1998) led to the detection of
X--ray pulsations with a period
of 15.83\,$\pm$\,0.02\,{\rm s}, later confirmed through pointed
\RXTE/PCA observations (Smith \& Takeshima 1998). A study of the lightcurve 
of RXTE/ASM has revealed an $\sim$\,80\,{\rm d} (or
160\,{\rm d}) X--ray modulation, interpreted as orbital
modulation (Campana et al. 1999).

The 1998 outburst was also monitored through a \BSAX\ campaign (Campana
et al. 1998) which
revealed a complex pulse profile (in 1--10\,{\rm keV}) and a complex
continuum spectrum, with an
iron line at 6.6\,{\rm keV} and an inferred column density of
$\sim$\,1.6\,$\times$\,$10^{22}$\,{\rm cm}$^{-2}$.  Using \RXTE\ data,
a cyclotron line has been
detected at $\sim$\,35\,{\rm keV} with a FWHM of $\sim$\,8\,{\rm keV}
(Heindl et al. 2001),
implying a magnetic field strenght of about 3.1(1+$z$)$\times$10$^{12}$
G, where $z$ is the
gravitational redshift of the scattering region.
\\
\indent In this paper we report on the identification of the optical/IR
counterpart of \src\ with
a $R$\,$\sim$\,$15$, Be spectral--type star. Based on \BSAX\
observations (circular uncertainty
region with a radius of 40\arcsec), we detected a reddened
($V$--$R$\,$=$\,1.6) star, showing in
the optical spectrum a strong H$\alpha$ emission line. The field of
\src\ has been subsequently
observed in the IR band. The candidate counterpart resulted as the
second brightest object in the
\BSAX\ error circle (H\,$\sim$\,12 mag), while the detection of Bracket
H$\gamma$ emission lines
in the IR spectra confirmed the Be spectral classification of the star.
A recently refined
\BSAX\ error circle (22\arcsec\ of radius) has finally confirmed the
identification ruling out the
possible link between the X--ray transient pulsar and the brightest IR
source included in the
previous \BSAX\ uncertainty region.
\begin{figure}[hpt]
\centerline{\psfig{figure=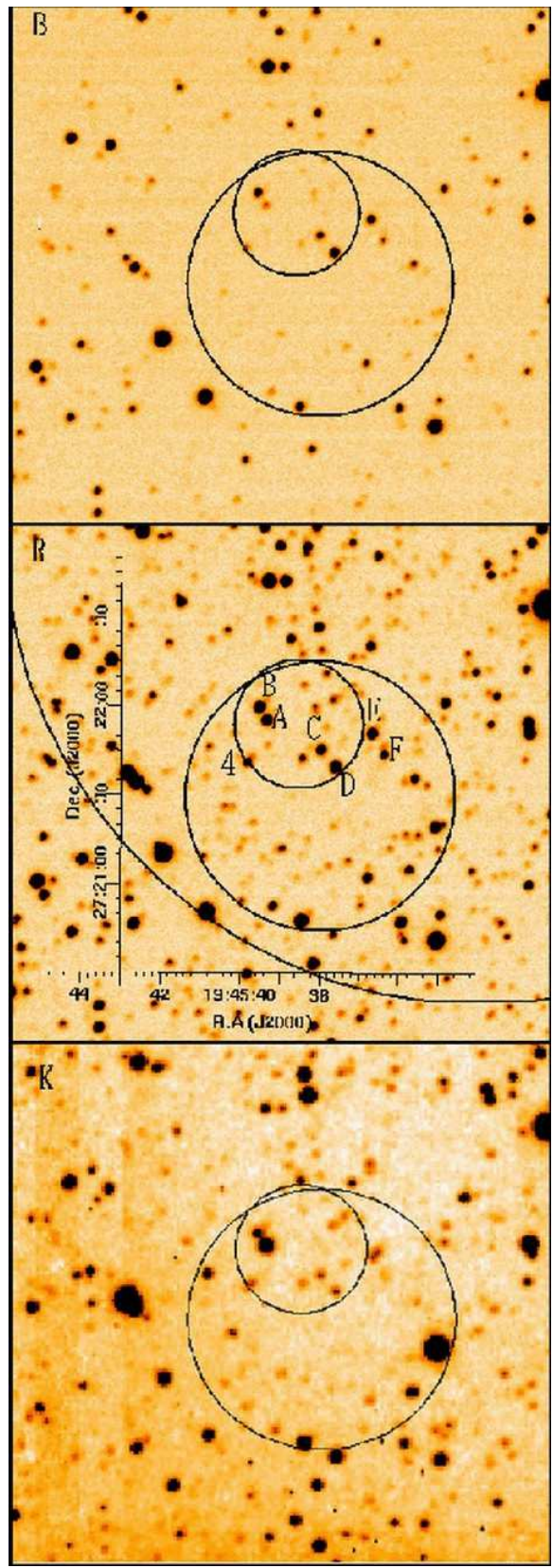,width=7.6cm} }
\caption{Optical $B, R$--filter (upper
panels) and IR $K$--filter (lower panel) images of the field of \src\
together with the X--ray position uncertainty circles derived from
\BSAX\ (inner circles) and \RXTE\ (larger circle in the
$R$--filter image) observations; R.A. ($h\, m\, s$) and Dec. ($^{\circ}$\, \arcmin\, \arcsec) 
coordinates are shown in the central panel (equinox J2000). The object marked with A 
is the proposed optical counterpart.}
\end{figure}

\section{The X--ray position}

The positional uncertainty region obtained with the \RXTE/PCA is a
circle with a radius of
2\arcmin.4 (90\% confidence level; Takeshima \& Chakra-\linebreak barty
1998); the latter was
reduced with \BSAX\ Narrow Field Instruments pointed observations to
40\arcsec\ (90\% confidence
level; Campana et al. 1998). The \BSAX\ X--ray position was
R.A.=19$^h$45$^m$38$^s$,
Dec.=+27\degr21\arcmin.5 (equinox 2000.0).  This position could
be refined thanks to a detailed
study performed on 73 relatively bright sources with known optical
position observed by \BSAX\
during its life (see also Fiore et al. 2001). The position uncertainty
was found to depend on several factors (intensity, off--axis angle,
roll angle, energy--dependent PSF, number of gyros and star--treakers,
etc.) some of which can be taken properly into account ending up with a
MECS error radius of 14\arcsec\ and 22\arcsec\ at 68\% and 90\%
confidence level, respectively. The first \BSAX\ position was than corrected for
these systematic uncertainties by performing a simple boresight
correction: the final \src\ position is R.A.= 19$^h$45$^m$38\fs5, Decl.=
+27\degr21\arcmin54\farcs3 (radius of 22\arcsec\ at 90\% confidence
level, equinox 2000.0).
\begin{table*}[hpt]
\begin{center}
\caption{Optical/IR observations carried out for the field of \src.}
\begin{tabular}{lcllccl}
\hline Telescope \& Instrument& Date & Exp. & Seeing & Range/Slit &
Candidate & Cand. A H$\alpha$
EW\\
                       &       &  (s) & (\arcsec) &(\AA/\arcsec) &  &
(\AA)\\
\hline 1.5\,m Cassini \& BFOSC & 1999 July 11 & 2400 & 2.5 &
4000--9000/2.5 & C,D &  ~~~----- \\
~~~~~~~~------  & 1999 July 12  & 150 & 1.9& V  & & ~~~----- \\
~~~~~~~~------ & ~~~~~"~~~~& 300 &
"&B &  &  ~~~----- \\ ~~~~~~~~------ & ~~~~~"~~~~& 100 & "&R & &
~~~----- \\ ~~~~~~~~------ &
~~~~~"~~~~& 60 & "&I &  &  ~~~----- \\ ~~~~~~~~------ & ~~~~~"~~~~&
1800 & " & 4000--9000/2.5& B,A
& --39\,$\pm$\,7 \\ ~~~~~~~~------ & 1999 July 13 & 2400 & 3.3 &
4000--9000/2.5 & E,F &   \\
~~~~~~~~------ & ~~~~~"~~~~& 1800 & " & " & B,A & --38\,$\pm$\,7 \\
~~~~~~~~------ & 1999 July 14
& 2400 & 2.0 & 4000--9000/2.0 & 4 & ~~~----- \\ ~~~~~~~~------ &
~~~~~"~~~~& 5400 & " & " & A &
--40\,$\pm$\,10 \\ ~~~~~~~~------ & 1999 July 15 & 2400 & 2.6 &
4000--9000/2.5 & A &
--40\,$\pm$\,12\\ ~~~~~~~~------ & 1999 July 16 & 4500 & 2.0 &
3300--5800/2.5 & A & ~~~----- \\
~~~~~~~~------ & 1999 July 17 & 2700 & 1.6 & " & A & ~~~----- \\
~~~~~~~~------ & 1999 July 19 & 2700 & 1.5 & 5800--8300/2.5 & A &
--42\,$\pm$\,3 \\ ~~~~~~~~------
& ~~~~~"~~~~& 2700 & " & 5800--8300/2.0 & A & --40\,$\pm$\,4 \\
~~~~~~~~------ & ~~~~~"~~~~& 1800
& " & 4000--9000/2.5 & A & --45\,$\pm$\,8 \\
 3.8\,m UKIRT \& CGS4 & 1999 July 24 & 300 & 1.5 &20400--22000/1.0  & A
&  ~~~----- \\
 1.1\,m AZT--24 \& SWIRCAM & 1999 July 27 & 60 & 2.1 & J &  & ~~~-----
\\
~~~~~~~~------ & 1999 July 27 & 60 & " & H &  & ~~~-----  \\
~~~~~~~~------ & 1999 July 27 & 60 &
" & K &  & ~~~-----  \\
 4.2\,m WHT \& ISIS & 2000 July 18 & 2400 & 0.9 & 3200--9000/1.0 & A  &
--41.0\,$\pm$\,0.5 \\
1.5\,m Cassini \& BFOSC  & 2000 July 31 & 3600 & 2.0 & 5000--9500/2.5 &
A & --40\,$\pm$\,8 \\
~~~~~~~~------ & 2000 August 28 & 1800 & 3.0 & 4000--9000/2.5 & A &
--33\,$\pm$\,5 \\
~~~~~~~~------ & 2001 June 18 & 3600 & 1.8  & 3600--8000/2.0 & A &
--40\,$\pm$\,5 \\ \hline
\end{tabular}
\label{tab:LogSpa}
\end{center}
\end{table*}

\section{Optical/IR observations}

We carried out observations in the optical and IR band of the \BSAX\
error circle in several
observing periods during 1999--2001 (see Table\,1) at the Loiano
observing station (Bologna Observatory), at Campo Imperatore (Observatory 
of Rome), at the United Kingdom Infrared Telescope (UKIRT, Mauna Kea, 
Hawaii) and at the William Herschel Telescope (WHT, La Palma, Spain).

\subsection{Imaging}

On July 1999 we performed imaging and photometry in the $B$ (300\,{\rm
s}), $V$ (150\,{\rm s}),
$R$ (100\,{\rm s}) and $I$ (60\,{\rm s}) filters with the 1.5-{\rm m}
Cassini telescope equipped
with the Bologna Faint Objects Spectrometer and Camera (BFOSC; Bregoli
et al. 1987; Merighi et al. 1994; 12\farcm2\,$\times$\,12\farcm2 field 
of view and 0\farcs58/pixel resolution, and a maximum resolution of 
$\sim$\,900 in the 3500--9000\AA\ range with a 1\farcs0 slit). Data 
reduction was performed using standard ESO--MIDAS procedures for 
bias subtraction and flat--field correction. Aperture and profile--fitting 
photometry for each stellar object in the image was
derived with the DAOPHOT\,II program (Stetson 1987).

\begin{table*}[htb]
\begin{center}
\caption{Optical and IR results for the objects in the \BSAX\ error
circle. Object A is the
proposed counterpart.}
\begin{tabular}{ccccccccccc}
\hline Object & $B$ & $V$ & $R$ &$I$ & $V-R$ & $J$ & $H$ & $K$ & $J-H$
\\
 & ($\pm$\,0.05) &  &($\pm$0.03)  & ($\pm$0.02) &  & ($\pm$0.1) &
($\pm$0.1) & ($\pm$0.1) &  \\ \hline
 A & 18.62 & 16.90$\pm$0.03 & 15.32 & 13.32 & 1.58 & 12.7 & 12.1 & 11.6
& 0.6  \\
 B & 17.29 & 15.53$\pm$0.03 & 15.13 & 13.74 & 0.94 & 13.9 & 13.3 & 13.2
& 0.6 \\
 C & 17.71 & 16.54$\pm$0.04 & 15.68 & 14.37 & 0.86 & 14.7 & 14.2 & 14.1
& 0.5 \\
 D & 16.71 & 15.71$\pm$0.03 & 14.96 & 13.76 & 0.75 & 14.3 & 13.9 & 14.0
& 0.4 \\
 E & 17.15 & 16.06$\pm$0.03 & 15.33 & 14.08 & 0.73 & 14.5 & 13.9 & 13.8
& 0.6 \\
 F & 18.89 & 17.33$\pm$0.04 & 16.50 & 15.16 & 0.83 & 15.7 & 15.2 & 15.1
& 0.5 \\
 4 & 18.94 & 17.56$\pm$0.03 & 16.66 & 15.22 & 0.90 & --& -- & -- & --
\\
\hline \\ \multicolumn{2}{l}{R.A.(J2000) = }&
\multicolumn{8}{l}{~~\,19$^{\rm h}$ 45$^{\rm m}$
39\fs3}\\ \multicolumn{2}{l}{Dec.(J2000) = } &
\multicolumn{8}{l}{+27$^{\circ}$ 21\arcmin\
~55\farcs4}\\ \multicolumn{2}{l}{Spectral type:} &
\multicolumn{8}{l}{~~\,B0--B1\,IV--V\,e}\\
\multicolumn{2}{l}{$E_{B-V}$:} &
\multicolumn{8}{l}{~~$1.9\,\rightarrow\,2.2$}\\
\multicolumn{2}{l}{Distance:} &
\multicolumn{8}{l}{~~$\sim$8--10\,kpc}\\ &&&&&&&\\ \hline
\end{tabular}
\end{center}
Note ---  Photometric uncertainties are at 90\% confidence level.
Position uncertainty is
0\farcs4.\\
\end{table*}
 We performed astrometry of the field using the USNO
catalog, resulting in a better than 0\farcs4 positional accuracy.  The
same field was observed in the IR $J$ (60\,{\rm s}), $H$ (60\,{\rm s}) 
and $K$ (60\,{\rm s}) filters on July 1999 with the 1.1-{\rm m} AZT--24 
telescope equipped with the Supernova Watchdogging IR Camera (SWIRCAM;
4\farcm4\,$\times$\,4\farcm4 field of view and 1\farcs04/pixel
resolution). Similar analysis procedures were applied to IR data.

Figure 1 shows the $B$, $R$ and $K$--filter images of the field
which includes the \src\ position: the 90\% position uncertainty circles 
as inferred by \RXTE\ (the largest one in the $R$--filter image) and 
\BSAX\ (the inner circles) are also shown.

Within the refined \BSAX\ X--ray positional error circle we detected
seven relatively bright (see Figure\,1) stars among which only one (labeled 
as A) was a reddened ($V$--$R$\,=\,$1.6$, all others had $V$--$R$\,$<$\,$1$) 
and relatively bright IR object ($H$\,=\,$12.1$).  In Table\,2 we
list results of optical/IR photometry for all seven stars.

\subsection{Optical Spectroscopy}

During July 1999 we also performed low--resolution ($\sim$\,18\AA)
spectroscopy of all relatively bright stars within the refined \BSAX\
circle (see Table\,1).  We applied standard corrections to spectra,
i.e. one--dimensional stellar and sky spectra extraction, then removed
cosmic rays and, when weather condition were satisfactory, we performed 
absolute calibration both for spectroscopy and photometry.
\begin{figure*}[htb]
\centerline{\psfig{figure=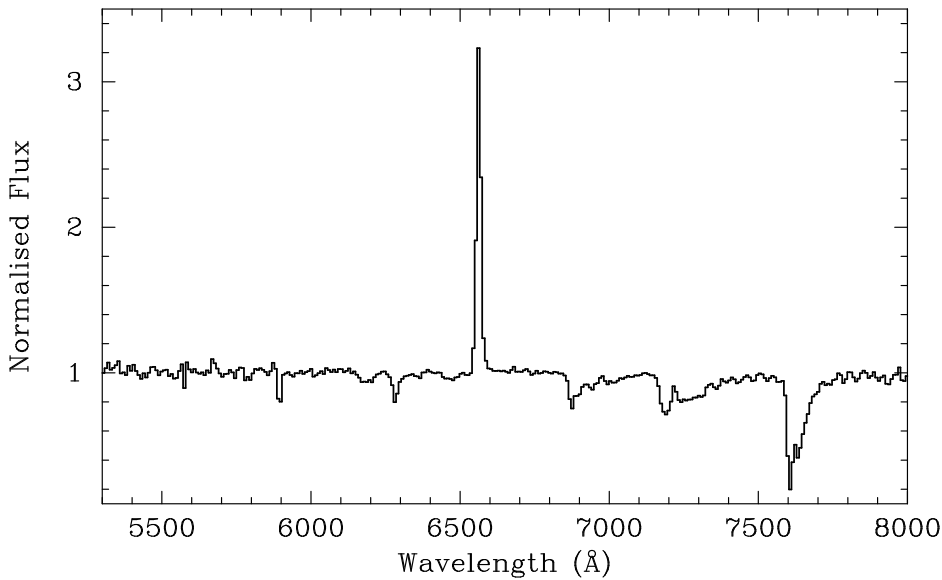,height=7.5cm,angle=-90} }
\caption{The 3600-s low resolution (18\AA) spectrum of \src\ 
candidate A (see Fig.\,1 middle panel) obtained on 2000 July 31 from the
1.5\,m Cassini telescope at Loiano (Bologna, Italy).}
\end{figure*}

All the spectra carried out in July 1999 in the 4000--9000\AA\
band (with different slit apertures) were rebinned to the largest 
resolution, 18.4\AA, and then summed to increase the S/N
ratio. None of the objects showed emission lines except star A.
For the latter source we detected a strong emission line (equivalent 
width, EW, of $\sim$\,--43\,$\pm$\,5\AA\ and Full Width Half Maximum, 
FWHM, comparable with the spectral resolution) with central wavelength
corresponding to that of H$\alpha$ (see e.g. Fig.\,2 for a
5000--9500\AA\ spectrum of star A obtained during a 3600\,{\rm s} observation
performed on 2000 July 31). Similarly we reduced, rebinned (9\AA\ of
resolution) and summed two spectra of star A in the band 5800--8300\AA\ 
to better characterize the H$\alpha$ emission line
(EW$\sim$\,--42\,$\pm$\,2\AA).

Spectra for star A were subsequently carried out routinely during
several 2000--2001 observational nights to study the possible presence of
variations of emission line parameters. The H$\alpha$ EWs
are consistent with being constant during our observations (see
Table\,1).
\begin{figure*}[htb]
\centerline{\psfig{figure=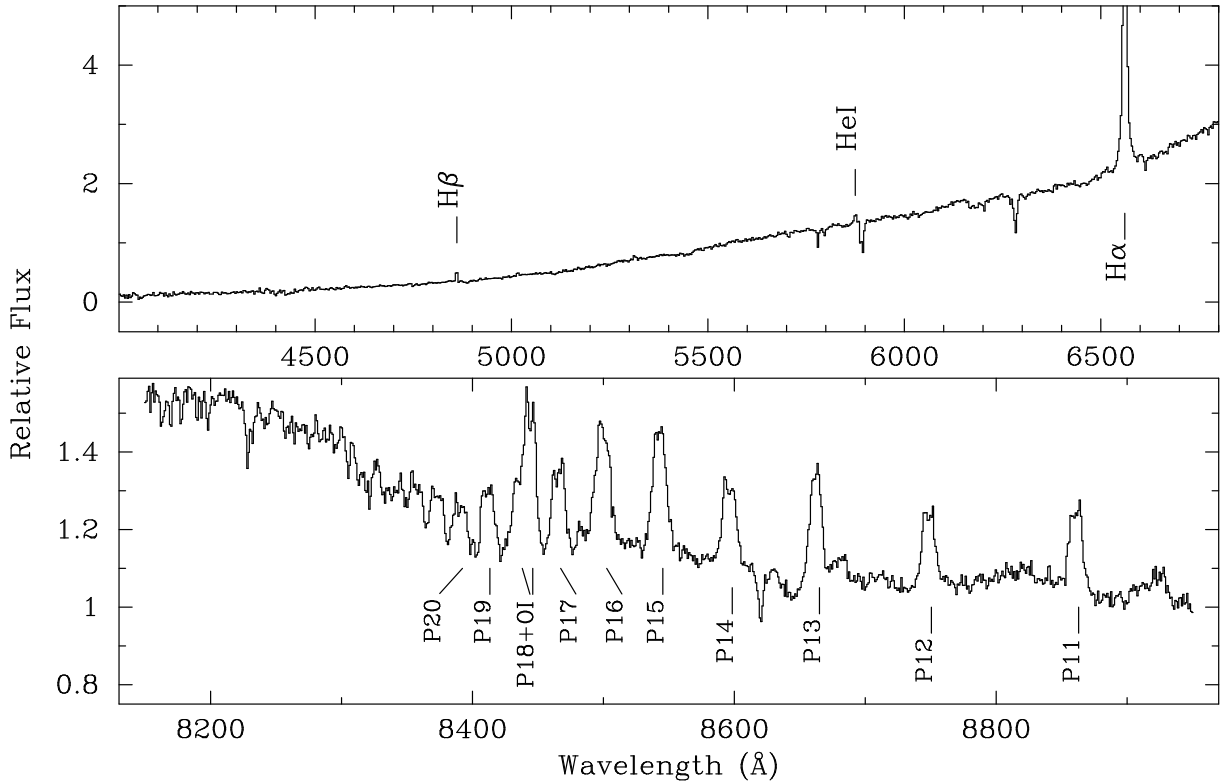,height=10.5cm,angle=-90} }
\caption{The 2400-s spectrum
(3.3\AA\ and 1.6\AA\ spectral resolution in the 4000--7000\AA\ and
8000--9000\AA\ range, respectively) of \src\ candidate A obtained on 
2000 July 18 from the 4.2\,m William Herschel Telescope at Roque de los 
Muchachos (La Palma, Spain). Balmer, Paschen and helium series lines are 
indicated.}
\end{figure*}
Owing to the faintness of the source shortwards of 5000\AA, no emission
lines have been detected in the blue part of spectra, with the
exception of a possible emission (EW\,$\sim$\,--6.0\,$\pm$\,0.3\AA) 
corresponding to H$\beta$.

Finally, on 2000 July 18 two 1200-s medium--resolution spectra
were taken with the Intermediate Dispersion Spectroscopic and Imaging 
System (ISIS) mounted on the 4.2\,m William Herschel
Telescope (WHT), located at the Observatorio del Roque de los
Muchachos, (La Palma, Spain). The blue arm was equipped with the R300B 
grating and the EEV\#10 CCD, which gives a nominal dispersion 
of $\sim$\,$0.9$\AA/pixel.  The resolution at H$\alpha$, estimated from 
the FWHM of arc lines, is $\sim$\,$3.3$\AA. The red arm was equipped with 
the R600R grating and the Tek4 CCD, which gives a nominal dispersion of 
$\sim$\,$0.8$\AA/pixel (the resolution is $\sim$\,$1.6$\AA\ at 
$\lambda$\,$\sim$\,$8500$\AA). The blue part of the spectrum is dominated 
by strong diffuse absorption bands (see Fig.\,3). The only feature 
clearly recognizable shortwards of H$\beta$ (which is clearly 
in emission with EW of $\sim$\,--2.0\,$\pm$\,0.5\AA) is the 
$\lambda$4430\AA\ DIB. An exact spectral classification is therefore 
impossible. However, based on the usual properties of Be
stars, we can strongly constrain the spectral type. The presence of
\ion{He}{i}~$\lambda$5875\AA\ strongly in emission is only observed in 
the earliest Be stars (earlier than B2).  Likewise, the presence of 
strong emission in the Paschen series (labelled with P in the 
lower panel of Fig.\,3) confirms that the object is earlier than B2
(Andrillat et al. 1988). As the lack of a detectable
\ion{He}{ii}~$\lambda$5412\AA\ absorption line imposes a spectral type 
later than O9.5, the object is constrained to lie in the B0-B1 range.
This is within the spectral range occupied by known Galactic and LMC
Be/X-ray binary counterparts, which have spectral types tightly 
concentrated around B0 (Negueruela \& Coe 2002).

\subsection{IR Spectroscopy}

\begin{figure}[htb]
\centerline{\psfig{figure=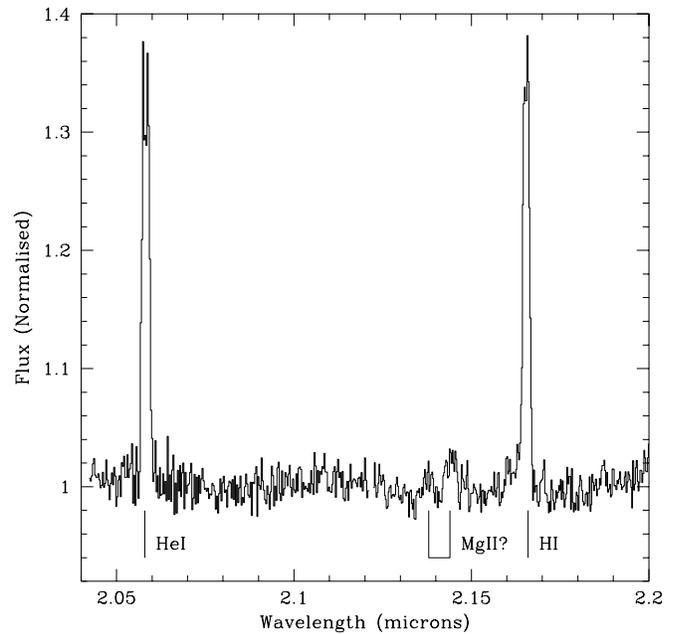,width=9cm} }
\caption{The 300-s IR
spectrum of \src\ candidate A obtained on 1999 July 24 from the UKIRT (Maunea Kea, Hawaii).}
\end{figure}
Spectroscopic IR observations of \src\ candidate were made with the United
Kingdom Infrared Telescope (UKIRT; Joint Astronomy Centre, Mauna Kea,Hawaii) 
on 1999 July 24, using the Cooled Grating Spectrometer (CGS4). Observations 
between 2.04--2.2~microns were made using the long focal length
camera plus the 150 line/\,{\rm mm} grating, giving a velocity
resolution of $\sim$\,50\,{\rm km/s}. 
 Initial data reduction was carried out at the telescope using the 
 {\footnotesize CGS4DR} software (Puxley et al. 1992).
This removes bad pixels, debiases, flat-fields, linearity corrects and
interleaves oversampled scan positions. The subsequent stages of data 
reduction, comprising of sky subtraction, extraction, derippling and 
wavelength calibration, were carried out using the Starlink-supported 
package {\footnotesize FIGARO}. Removal of telluric features was accomplished 
via the procedure described by Clark \& Steele (2000).

The spectrum (see Fig.\,4) shows strong emission \linebreak from He\,{\sc i} 
2.058\,$\mu$\,{\rm m} (EW\,=\,--8.1\,$\pm$\,0.4\AA) and H\,{\sc i} Bracket $\gamma$
(EW\,=\, --8.0\,$\pm$\,0.4\AA). FWHMs for the two lines are identical 
within the errors on the measurements
and yield a mean value of 300\,$\pm$\,50\,{\rm kms}$^{-1}$. There is
some evidence for weak emission in the Mg\,{\sc ii} 2.138/2.144\,$\mu$m doublet
(EW\,=\,--0.6\,$\pm$\,0.4\AA\ and --1.4\,$\pm$\,0.4\AA, respectively), 
although further higher S/N observations are required to
confirm this. If the emission features are real the ratio of the EW of
both lines suggests that the lines are optically thin, and are expected 
to be excited via Ly$\beta$ fluorescence. No evidence for emission from 
Fe\,{\sc ii} 2.189\,$\mu$m or He\,{\sc i} 2.112/3\,$\mu$m emission or
absorption was found.

Comparison to the $K$ band spectra of classical Be stars presented by
Clark \& Steele (2000) suggests that \src\ candidate A
can be classified as a Group 1 object consistent with it being of early
(O9--B3) spectral type (given the paucity of photospheric features in
the near IR it is not possible to classify a Be star to greater accuracy 
than $\pm$\,2 spectral types or provide a luminosity class) in agreement 
with the optical spectroscopic findings. We find no anomalous
emission from higher excitation species to suggest that the star has a
compact companion; in line with the trend displayed by other Be/X-ray 
binaries for which K band spectra have been obtained which also
have spectra indistinguishable from those of isolated early spectral
type classical Be stars (e.g. Clark et al. 1999).

\section{Discussion and Conclusion}

We have presented the X--ray, optical and IR observational data of the
field containing the error circle of \src; these observations led 
to the likely identification of the optical counterpart of
this 15.8-s transient X--ray pulsar. Optical/IR photometric
measurements are consistent with a relatively reddened and distant 
blue star, similar to known optical/IR counterparts of BeXRBs (see
Nelson et al. 1993 \& 1995; Israel et al. 2000a; Israel et al. 2000b;
Covino et al. 2000; Covino et al. 2001, Israel et al. 2001; Chakrabarty 
et al. 2002).

Owing to the uncertainty in the spectral classification, the
measurement of the distance from optical data becomes difficult; 
it is however possible to extract some information from our
optical and IR data. The observed $V$--$R$ color for \src\ is
$\sim$\,1.6, while the intrinsic one
should be $\sim\,-0.14 \rightarrow -0.11$ (assuming a main sequence or
sub--giant star with spectral class in the B0--B1\,III--V range; 
Wegner 1994), so the reddening should amount to
$\sim\,1.7 \rightarrow 1.8$. Assuming a standard reddening law
(Fitzpatrick 1999) this converts to $A_{R}\sim\,5.0 \rightarrow 5.1$, 
$A_{V}\sim\,6.7 \rightarrow 6.8$ and $E_{B-V}\sim 2.0\rightarrow2.2$ 
(regardless of the distribution of the medium responsible 
for the reddening along the line of sight and near the source, i.e. 
assuming that the reddening due to circumstellar disk
is negligible).

However from the X--ray spectral data (Campana et al. 1998) a $N_H$ of
$\sim$\,1.6\,$\times10^{22}$\,{\rm cm}$^{-2}$ was inferred,
corresponding to an $E_{B-V}$\,$\sim$\,2.9 
(Predehl \& Schmitt 1995), which is slightly more than what obtained from
optical data.  Considering the total Galaxy $N_H$ column in the
direction of \src\ (Dickey \& Lockman  1990), $<$1$\times10^{22}$ cm$^{-2}$, 
a value of $E_{B-V}\sim 1.9 \rightarrow 2.1$ was
inferred, which comfortably overlaps the value inferred from the
optical data. This result suggests that at least part of the inferred X--ray 
$N_H$ is local to the system and obscures the
neutron star during outbursts.  A good agreement with the $B$ to $I$
measurements is obtained for a B0--1IV--V star at a distance of about 
8--10\,kpc and with $E_{B-V}\sim\,2.$ (see Table\,2). We
can reasonably discard the possibility of a luminosity class III which
would imply an $E_{B-V}$\,$>$\,$2.3$ and a large IR--deficiency. For 
a reference B0--1IV--V star ($M_{V}\sim-3.5$) at a distance of 8--10\,kpc 
and based on the IR photometry we infer an excess of
$\sim$0.4, 0.5 and 1.4 magnitudes in $J$, $H$ and $K$ filters
respectively, suggesting the presence of a circumstellar envelope. Finally 
we note that in the direction of \src\ there are two spiral arms of our
Galaxy located at $\sim$4\,kpc and $\sim$9\,kpc away from the Earth 
(Perseus and Cygnus arms; Turner 1980, Taylor \& Cordes 1993,
Vall\'{e}e 2002), so \src\ could belong to the second one.
Moreover this distance estimation is very similar to that found for the
nearby source KS 1947+300 (Negueruela et al. 2002), 
which could lie on the same arm.

For a distance of 8--10\,kpc and a 1--10\,keV flux of
$\sim$\,1\,$\times$\,10$^{-9}$\,erg~s$^{-1}$~cm$^{-2}$ at the peak of
the 1998 outburst (Takeshima \& Chakrabarty 1998) we obtain an X--ray 
luminosity of L$_{\rm X}$\,(1--10\,keV)\,$\simeq$\,8--20\,$\times$
\,10$^{36}$\,erg~s$^{-1}$.
Such a luminosity is a typical value shown by X--ray pulsars in binary 
systems during Type I outbursts (Stella et al. 1986; Negueruela 1998) 
occurring close to the time of periastron passage and with a periodic 
recurrence at the orbital period of the system.

Based on both X--ray and optical/IR observations we identify 
the likely optical/IR counterpart of \src.
Moreover, photometric and spectroscopic optical/IR data allows us to
conclude that the proposed optical counterpart is most likely a B0--1\,V--IVe 
star at a distance of about $\sim$8--10\,kpc. A
more accurate distance and spectral classification would require more
detailed optical/IR spectroscopic observations.

\begin{acknowledgements}

This work is partially based on observations carried out at Loiano
observing station (Bologna Observatory).
 The WHT is operated on the island of La Palma by the Isaac Newton
Group in the Spanish Observatorio del Roque de los Muchachos of the Instituto
de Astrof\'{\i}sica de
Canarias. We thank the service programme for performing the
observations. The United Kingdom
Infrared Telescope is operated by the Joint Astronomy Centre on behalf
of the U.K. Particle
Physics and Astronomy Research Council. JSC acknowledges funding
support from PPARC. 
Moreover authors would like to thank the referee, Dr. L.M. Kuiper, for useful advices.
\end{acknowledgements}

\newpage

\end{document}